\documentclass[12pt]{article}
\usepackage{lmodern}
\usepackage[english]{babel}
\usepackage{pdflscape, pslatex} 
\usepackage{epstopdf}
\usepackage{float}
\usepackage{bm}
\usepackage[authoryear]{natbib}
\usepackage{longtable}
\usepackage{times}
\usepackage{adjustbox} 
\usepackage{rotating}
\usepackage{url}
\usepackage{setspace}
\usepackage{listings} 
\usepackage{array}
\usepackage{multirow, multicol} 
\usepackage{tabulary,ctable,tabularx}
\usepackage{booktabs}
\usepackage{graphicx}
\usepackage{appendix}
\usepackage{enumerate}
\usepackage{times}
\usepackage{subfig}
\usepackage{amsmath,amsfonts,amssymb,amsthm,titling}

\usepackage[a4paper,width=160mm,top=30mm,bottom=30mm]{geometry}

\usepackage{authblk}

\title{Assessing the Performance of the Discrete Generalised Pareto Distribution in Modelling Non-Life Insurance Claims}

\author{S. K-B. Dzidzornu and R. Minkah\textsuperscript{}\thanks{\noindent\textsuperscript{}Correspondence: \textit {rminkah@ug.edu.gh}\\
	The second named author (R. Minkah) would like to thank the University of Ghana Building a New Generation of Academics in Africa (BANGA-Africa) Project with funding from Carnegie Corporation of New York for providing support through its write-shop program for the write-up of this paper. }\\Department of Statistics and Actuarial Science,\\
School of Physical and Mathematical Sciences\\College of Basic and Applied Sciences\\ University of Ghana, Ghana\\
}
\date{}

\begin{document}
	\def\spacingset#1{\renewcommand{\baselinestretch}%
		{#1}\small\normalsize} \spacingset{1}
\maketitle
\begin{abstract}
	In this paper, non-life insurance claims were modelled under the three parameter discrete generalised Pareto distribution. Data from the National Insurance Commission of Ghana on reported and settled claims were considered for the period 2012-2016. The maximum likelihood estimation principle was adopted in fitting the discrete Pareto distribution to the yearly and aggregated data. The estimation involved two steps. Firstly, the $\mu$ and $(\mu+1)$ frequency method of \citet{Prieto2014} was modified to suit the characteristics of the data under study. Secondly, a bootstrap algorithm was implemented to obtain the standard errors of the estimators of the parameters of the discrete generalised Pareto distribution. The performance of the discrete generalised Pareto distribution is compared to the negative binomial distribution in modelling the non-life insurance claims data using the information criteria of Akaike and Bayesian. The results show that the discrete generalised Pareto distribution provides a better fit to the non-life claims data.

\vspace{0.5em}
\end{abstract}

	\section{Introduction}\label{sec1}

Non-life or General Insurance involves the provision of financial loss protection against risks on interests other than life, such as buildings, vehicles, machinery and equipment. Conditioned on periodic payments or one-off advance of a predetermined amount, called premium, non-life policies are designed to provide coverage against the occurrence of the insured probabilistic events, for individuals, private organisations, and public institutions. The payments effected in response to occurrences of such events are termed as insurance claims \citep{Wuthrich2019}. The non-life insurance claims process is characterised by two features: claims frequency or count, and claims severity or size. As noted by \citet{Renshaw1994} and \citet{Ozgurel2005}, the underlying expectations of claims frequency and severity, quantified as a product, are foremost considerations in the computation of pure or risk premiums.

The main objective of this paper is to illustrate that the discrete generalised Pareto (DGP) distribution can be employed to model the counts of non-life insurance claims, collated by an insurance regulatory authority from a licensed class of insurers. In the absence of suitable actuarial models, non-life insurers largely encounter difficulties in conducting evidence-based assessment of risks insured, often resulting in the mis-computation of premiums, and inability to settle claims when due. In response, developing probability models that describe claims frequencies offer a distributional framework for evaluating risks, to facilitate premium setting and liquidity reserving by non-life insurance service providers.

A random variable $X$ that follows the discrete generalised Pareto distribution with shape ($\alpha$), scale ($\lambda$), and location ($\mu$) parameters is denoted by $X\sim DGP (\alpha,\lambda,\mu).$ This is a parametric model is obtained by discretising the continuous generalised Pareto distribution, introduced by \citet{Pickands1975} and it is particularly noted for tail modelling properties. The discrete generalised Pareto distribution assumes varying forms based on omission of one or two of its parameters. For instance, if the location parameter, $\mu=0$, the discrete generalised Pareto distribution transforms into the Discrete Lomax distribution, $DLo (\alpha,\lambda).$ 

A fair amount of research has demonstrated the application of probabilistic models to the study of non-life insurance claims. Among the relevant literature, selected studies have principally explored the subject with reference to standard probability distributions, from the outlook of randomised spatial effects  \citep{Gschlossl2007}, collective risk simulation \citep{Pacakova2011}, and incorporation of covariates (Renshaw, 1994). In this paper, we attempt to contribute to the gap by providing a method of fitting the discrete generalised Pareto distribution to non-life claims data from the National Insurance Commission in Ghana. Further, to illustrate the performance of the discrete generalised Pareto distribution, we compare it to the negative binomial distribution, similar to the work of \citep{Prieto2014}. The results show that the discrete generalised Pareto distribution is appropriate in modelling the count frequencies of non-life insurance claims.

The study contributes to the field of claims modelling in three folds. First, it proposes a count-based data scheme for analysis of non-life insurance claim frequencies, to enhance the precision of statistical models. Second, a modification is made to the $\mu$ and $\mu+ 1$  frequency method of \citet{Prieto2014}, to obtain initial estimators of the discrete generalised Pareto distribution in claims data characterised by varying discrete observational intervals. Third, the algorithm implemented under the estimation of the parameters of the discrete generalised Pareto distribution offers a resource to researchers, on performance analysis in future statistical and/or actuarial work.


The rest of the paper is organised as follows. In Section \ref{sec3}, the methodology is presented including the maximum likelihood estimation of the parameters of the distributions and model selection criteria. Section \ref{sec4} presents the data and the arrangements needed to put the data into a form necessary for the model fitting. Lastly, in Section \ref{sec5}, we present concluding remarks. 
\section{Methodology}\label{sec3}
This section presents the systematic approach followed to model the reported and settled claims datasets. Specifically, the section entails the description of the probability distributions, parameter estimation and model selection criteria. 

The parameter estimation method called the Maximum Likelihood Estimation (MLE) technique, would be used in fitting the models. Consider the case where a random variable is available from a population with a known probability distribution except for its parameter $\bm{\theta}\in \mathbb{R}^d.$ The maximum likelihood principle suggests that the criterion for making the selection should be the probability (or likelihood) with which a particular distribution can produce the given sample. The value of $\bm{\theta}$ for that distribution is the maximum likelihood estimate of the unknown $\bm{\theta}.$

Suppose $\bm{x}=\{x_1, ~x_2,~...,~ x_n\}'$ is an independent random sample of size, $n,$ from a distribution with dependence on one or more unknown parameters $\bm{\theta}=\{\theta_1, \theta_2,..., \theta_d\}'$. Let $f(x_i; \bm{\theta})$ be the probability density (or mass) function of $x_i,$ with $\bm{\theta}$ restricted to a given parameter space $\Omega\in\mathbb{R}^d.$ The likelihood function of the sample is given by
\[L(\bm{\theta}; \bm{x})=\prod_{i=1}^{n} f(x_i).\]
The maximum likelihood estimator, $\hat{\bm{\theta}},$ of $\bm{\theta}$ is the solution to the equation 
\[ \frac{\delta}{\delta\bm{\theta} }L(\bm{\theta}; \bm{x})=0.\]
Usually, the $L(\bm{\theta}; \bm{x})$ may involve exponentials, and hence,  $\ln L(\bm{\theta}; \bm{x})$ is maximised. Since the logarithm of a function increases or decreases with the function, the maximiser of $L(\bm{\theta}; \bm{x})$ also maximises $\ln L(\bm{\theta}; \bm{x}).$

\subsection{Negative Binomial Distribution} 

The negative binomial is a discrete probability distribution which characterises the number of successes in a sequence of independent and identically distributed Bernoulli trials before a specified number of failures (denoted r) occurs.
Suppose a sequence of Bernoulli trials is observed. By definition, a turn of each trial yields two possible outcomes, success or failure, with respective probabilities of occurrence denoted by $p$ and $1-p$. Also, the trials are independent and $p$ remains constant for each trial. If $X$ represents the number of trials (or failures) prior to the $r$-th success, then $X$ follows a negative binomial distribution with probability mass function:

\begin{equation}\label{ngbin}
P(X=x) ={x - 1 \choose r - 1}(1 - p)^{x - r} \, p^r ~\mbox{for} ~ x = r, r + 1, r + 2, \cdots
\end{equation}

The geometric distribution is a special case of the negative binomial, where the Bernoulli trial discontinues at first failure, $r = 1.$ Since the negative binomial distribution may be represented by alternative parametrisations, three factors inform distinctions: starting point of the support - whether at $x = 0$ or $x = r;$ definition of $p$ - whether it represents the probability of success or failure, and interpretation of $r$ - whether it denotes the number of success or failure \citep{Degroot2012}. 

Given $N$ independent and identically distributed claims count observations, $(k_1, \cdots, k_N)$, the likelihood function can be expressed as:

\begin{equation}\label{eq1}
L(r,p) = \prod_{i=1}^{N}P(X=k_i;r,p). 
\end{equation}
Substituting (\ref{ngbin}) into (\ref{eq1}) and taking the logarithm results in the log-likelihood function given by

\begin{equation}\label{eq2}
\ell(r,p) = \sum_{i=1}^{N}\ln(\Gamma(k_i + r)) - \sum_{i=1}^{N}\ln(k_i!) - N\ln(\Gamma(r)) + \sum_{i=1}^{N}k_i\ln(p) + Nr\ln(1 - p).
\end{equation}
To maximise equation (\ref{eq2}), the partial derivative with respect to $r$ and $p$ are set to zero (0),
\begin{equation}\label{eq3}
\frac{\partial l(r,p)}{\partial p} = \left[\sum_{i=1}^{N}k_i\frac{1}{p}\right] - Nr\frac{1}{1 - p} = 0
\end{equation}

and

\begin{equation}\label{eq4}
\frac{\partial l(r,p)}{\partial r} = \left[\sum_{i=1}^{N}\psi(k_i + r)\right] - N\psi(r) + N\ln(1 - p) = 0.
\end{equation}
Here the digamma function $\psi(k)=\Gamma'(k)/\Gamma(k).$ Furthermore, solving for $p$ in equation (\ref{eq3}) produces:
\begin{equation}
\hat{p} = \frac{\sum_{i=1}^{N}k_i}{Nr + \sum_{i=1}^{N}k_i}.
\end{equation}
Finally, substituting $p$ in (\ref{eq4}) yields,

\begin{equation}\label{eq5}
\frac{\partial l(r,p)}{\partial r} = \left[\sum_{i=1}^{N}\psi(k_i + r)\right] - N\psi(r) + N\ln\left(\frac{r}{r + \sum_{i=1}^{N}k_i/N}\right) = 0
\end{equation}
The form of (\ref{eq5}) suggests that a closed form solution for $r$ may not be obtained analytically. Therefore, numerical methods can be used in order to obtain estimators of $r$ and $p.$ For example, in \citet{Rcore}, the function  \textit{fitdistr} in the \textit{MASS} package provides a routine for estimating the parameters of the negative binomial distribution.

\subsection{Discrete Generalised Pareto Distribution}
The discrete generalised Pareto  distribution arises from a discretisation of the continuous generalised Pareto distribution. To provide a basis for the discussion on the discrete generalised Pareto, the Pareto Type-I and generalised Pareto are given as
\begin{equation}\label{eq6}
D(x) = 1 - \left(\frac{b}{x}\right)^a,~x\geq b
\end{equation}
and

\begin{equation}\label{eq7}
G(x) = 1 - \left[1 + \frac{\alpha(x - \mu)}{\lambda}\right]^\frac{-1}{\alpha},~1 + \alpha(x - \mu)/\lambda > 0,  x > \mu,~ \mbox{and}~\lambda > 0,
\end{equation}
respectively. The generalised Pareto distribution is noted for its ability to model tails of distribution functions \citep[See e.g.][]{Davison1990,Beirlant2004}. Upon discretisation of the generalised Pareto distribution, the resultant  discrete generalised Pareto distribution inherits the prior continuous properties in forms adapted to the discrete probability space.

From the stated generalised Pareto distribution (\ref{eq7}), the probability mass function of the  discrete generalised Pareto can be formally deduced. First, consider the cumulative distribution function  of the discrete generalised Pareto expressed as:

\begin{equation}\label{eq8}
F(x) = P(X \leq x) = 1 - [1 + \lambda(x - \mu + 1)]^{-\alpha},~x = \mu, \mu + 1, \cdots,
\end{equation}
where $~\alpha, \lambda,\mu > 0$ and $F(x) = 0$ if $x < \mu.$ 

Also, \citet{Krishna2009} addressed the discretisation of a continuous model by observing unit groupings on failure time axis. The authors reasoned that for a continuous failure time $X$, with survival function $S(x) = P[X> x]$ and time groupings of intervals $dX = \lfloor{X}\rfloor$, the discrete observed variable, $dX$, would have the probability mass function, 

\begin{equation}\label{eq9}
P(dX = x) = P(x \leq X < x + 1) = S(x) - S(x + 1),~ x = 0, 1, 2, \cdots.
\end{equation}
Next, consider a standard generalisation for survival function from \citet{Xekalaki1983},

\begin{equation}\label{eq10}
S(x) = P(X \ge x) = 1 - F(x - 1),
\end{equation}
then, employing equations (\ref{eq8}) and (\ref{eq10}), the survival function of the discrete generalised Pareto distribution is given

\begin{equation}\label{eq11}
S(x) = [1 + \lambda(x - \mu)]^{-\alpha},~x = \mu, \mu + 1, \cdots.
\end{equation} 

Finally, evaluating equations (\ref{eq9}) and (\ref{eq11}) simultaneously result in the discrete generalised Pareto distribution,

\begin{equation}\label{eq12}
f(x)= \left[1+\lambda(x-\mu)\right]^{-\alpha} - \left[1 + \lambda(x-\mu+1)\right]^{-\alpha}, ~x = \mu, \mu+1, \cdots.
\end{equation}



Suppose $x_1, \cdots, x_n$ is a sample of size $n$ from a   discrete generalised Pareto distribution. The parameters $\alpha$ and $\lambda$ are estimated on the assumption that $\mu$ is known, since $\hat{\mu} = x_{min} \leq x_i, \forall i.$ Adopting the $\mu$ and $(\mu+1)$ frequencies method of \citet{Prieto2014}, the initial values, $(\alpha_0, \lambda_0, \mu_0)$ can be obtained and used as seed estimators in the subsequent maximum likelihood operation. Thus, the relative frequencies of $x = \mu$, and $x = (\mu+1)$, respectively denoted by $\hat{f}_{\mu}$ and $\hat{f}_{\mu+1}$, are calculated from the sample data. Analogously, $\alpha$ and $\lambda$ are determined by substituting $x = \mu$ and $x = (\mu+1)$ into the discrete generalised Pareto probability mass function in (\ref{eq12}) and equating the expressions to their respective $\hat{f}_{\mu}$ and $\hat{f}_{\mu+1}$ values.  

However, the $\mu$ and $(\mu+1)$ frequencies method assumes that the count data used is observed in increasing steps of 1. However, in real-life situations, such as the data presented in Section \ref{Data},  may exhibit variation of intervals other than 1. Therefore, applying the method strictly on the count data results in generating several $\mu+1 = 0,$ and hence, leading to $\hat{f}_{\mu+1} = 0.$ 

In that regard, proceeding with the computations with zero (0) relative frequencies, will result in a loss of essential frequency information in the dataset. As a result, we provide a modification of the method as $\mu$ and $\mu+\epsilon$ frequency methods, where $\epsilon > 0$ and ($\mu+\epsilon$) is the smallest observation larger than the minimum, $\mu.$ Therefore, the estimators of $\alpha$ and $\lambda$ are obtained by solving the resulting expressions,  

\begin{equation}\label{eq13}
\hat{f}_{\mu} = 1 - [1 + \lambda]^{-\alpha}
\end{equation}
and 
\begin{equation}\label{eq14}
\hat{f}_{\mu + \epsilon} = [1 + \lambda]^{-\alpha} - [1 + 2\lambda]^{-\alpha}
\end{equation}
simultaneously. The expression in (\ref{eq15}) results after $\alpha$ is eliminated from equations (\ref{eq13}) and (\ref{eq14}),

\begin{equation}\label{eq15}
\frac{\ln(1 + 2\lambda)}{\ln(1 + \lambda)} = \frac{\ln(1 - \hat{f}_{\mu} - \hat{f}_{\mu+\epsilon})}{\ln(1 - \hat{f}_{\mu})}. 
\end{equation}
Following from this, (\ref{eq16}) is obtained by appropriate substitutions into (\ref{eq15}),

\begin{equation}\label{eq16}
\hat{\alpha} = -\frac{\ln(1 - \hat{f}_{\mu})}{\ln(1 + \hat{\lambda})}.
\end{equation}

Next, the maximum likelihood estimation method is employed to obtain estimators of the parameters of the discrete generalised Pareto. The log-likelihood function is constructed as,

\begin{equation}\label{eq17}
\ln\ell(\lambda, \alpha) = \sum_{i=1}^{n}\ln f(x_i) = \sum_{i=1}^{n}\ln\left[(1 + \lambda(x_i - \mu))^{-\alpha} - (1 + \lambda(x_i - \mu + 1))^{-\alpha}\right],
\end{equation}
where $f(x_i)$ refers to the probability mass function specified in (\ref{eq12}). Partial derivatives of (\ref{eq17}) are taken with respect to $\alpha$ and $\lambda$, and set to zero (0) to obtain normal equations,

\begin{equation}\label{eq18}
\frac{\partial{\ln\ell}}{\partial{\alpha}} = \sum_{i=1}^{n}\frac{\ln[1 + \lambda(x_i - \mu + 1)]}{[1 + \lambda(x_i - \mu + 1)]^{\alpha}[1 + \lambda(x_i - \mu)]^{-\alpha} - 1} - \sum_{i=1}^{n}\frac{\ln[1 + \lambda(x_i - \mu)]}{1 - [1 + \lambda(x_i - \mu)]^{\alpha}[1 + \lambda(x_i - \mu + 1)]^{-\alpha}} = 0
\end{equation}

\[\frac{\partial{\ln\ell}}{\partial{\lambda}} = \sum_{i=1}^{n}\frac{\alpha(x_i - \mu + 1)}{[1 + \lambda(x_i - \mu + 1)]^{\alpha+1}[1 + \lambda(x_i - \mu)]^{-\alpha} - [1 + \lambda(x_i - \mu + 1)]}\]

\begin{equation}\label{eq19}
- \sum_{i=1}^{n}\frac{\alpha(x_i - \mu)}{[1 + \lambda(x_i - \mu)] - [1 + \lambda(x_i - \mu)]^{\alpha+1}[1 + \lambda(x_i - \mu + 1)]^{-\alpha}} = 0
\end{equation}

To proceed with estimation of the parameters of the discrete generalised Pareto distribution, a function in R was written to perform the following this algorithm:
\begin{enumerate}
	\item[A1.] Specify the log-likelihood function (\ref{eq17}) based on the discrete generalised Pareto probability mass function, (\ref{eq12}). The log-likelihood function is set to return a negation of the log-likelihood value, since the R \textit{optim} function is a minimiser. In effect, minimising the negated log-likelihood function at the initial estimates, produces the equivalent of maximising the log-likelihood function.
	
	\item[A2.] Optimise the log-likelihood functions in (\ref{eq18}) and (\ref{eq19}) at the seed values, by Simulated Annealing (SANN), a variant of \citet{Belisle1992} technique.
	
	\item[A3.] Extract the estimated parameters, $\alpha$ and $\lambda$ from the output generated in A2 and compute the standard errors of the estimators using bootstrap resampling of \citet{Efron1993}.
\end{enumerate}

\subsection{Model Selection Criteria} 
The \citet{Akaike1974} and  Bayesian \citet{Schwarz1978} Information Criteria, usually denoted AIC and BIC respectively, form the basis for selecting the suitable model. The AIC and BIC are stated as follows:

\begin{equation}\label{eq20}
AIC = -2\ln L + 2d
\end{equation}
and
\begin{equation}\label{eq21}
BIC = -2\ln L + d\ln(n)
\end{equation}

In comparison with AIC, BIC addresses the issue of overfitting with a factor, $\ln(n)$, thereby placing a higher penalty for model complexity \citep{Dziak2019}. In statistical decision making, a candidate model with minimum AIC and/or BIC values is selected.

\section{Data and Model Fitting}\label{sec4}

\subsection{Data} \label{Data}
The study employs secondary data on non-life claims from the National Insurance Commission (NIC) of Ghana for the five-year period, 2012 to 2016. The historical data covers insurance claims of twenty-nine (29) non-life service providers. For each fiscal year, the dataset indicates total number of claims administered under each of the five (5) business classes of non-life insurance in Ghana. The classes are: Fire, Burglary, and Property Damage; Accident; Marine and Aviation; Motor; and General Liability. The claims data is organised into three (3) categories: Incurred But Not Reported (IBNR), Reported But Not Settled (RBNS), and Settled But Outstanding (SEBO) each bearing the standard actuarial definitions.

However, since IBNR is necessarily an estimate, the study focuses on RBNS and SEBO, hereinafter referred to as reported and settled claims respectively. Overall, the data consists of 3,878,355 non-life insurance policies, generating 39,563 reported claims, of which 5,210 claims were settled within the period. 

\begin{figure}
	\centering
	\includegraphics[width=0.7\linewidth]{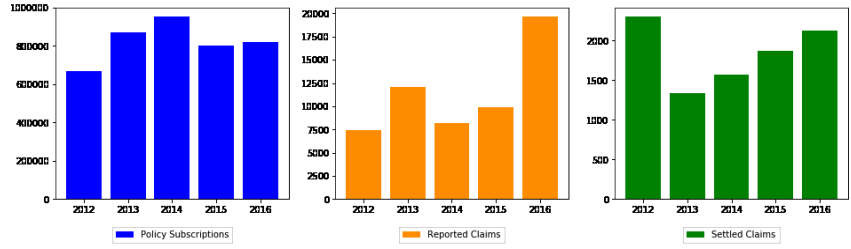}
	\caption{Claims Data: Number of non-life policies recorded in the period 2012-2016}
	\label{claims1}
\end{figure}

Figure \ref{claims1} provides an overview of the annual aggregates for policy subscriptions, reported claims, and settled claims. Although policy subscriptions have seen a decrease from 2014 onwards, the number of reported and settled claims have been increasing within the period. This observation shows some evidence of potential liquidity challenges for the non-life insurers, if the trend persists into the future.

Following \citet{Prieto2014}, the dataset is organised in such a way as to enable the fit of the discrete distributions to the frequency of occurrence of reported and settled claims. Tables \ref{tab1} and \ref{tab2} present descriptive statistics on the reported and settled claims datasets respectively. For each year of the period considered, the skewness indicates the extent of symmetry and shows that there is positive skewness for the distribution of count of claims. Also, among the reported claims, the fiscal year 2016 recorded some unusually large values culminating in its large kurtosis value. Similar results can be found in 2013 for the settled claims data.

\begin{table}[htp!]
	\centering
	\caption{Descriptive statistics of reported claims}
	\label{tab1}
	\begin{tabular}{lcccccc}
		\toprule
		Year&Mean&Std. Dev.&Skewness&Kurtosis&Min&Max\\
		\hline
		2012&255.86&329.95&1.28&0.71&0&1187\\
		2013&414.9&620.11&1.58&1.12&0&2053\\
		2014&281.9&350.03&1.57&1.47&0&1335\\
		2015&341.72&503.77&2.12&4.07&0&2151\\
		2016&676.59&1839.98&4.34&18.81&2&9902\\
		\bottomrule
	\end{tabular}
\end{table}

\begin{table}[htp!]
	\centering
	\caption{Descriptive statistics of settled claims}
	\label{tab2}
	\begin{tabular}{lcccccc}
		\toprule
		Year&Mean&Std. Dev.&Skewness&Kurtosis&Min&Max\\
		\hline
		2012&79.21&123.42&1.85&2.34&0&452\\
		2013&46.24&63.22&2.52&7.35&0&307\\
		2014&54.03&65.38&2.24&5.42&0&307\\
		2015&64.31&90.59&1.93&3.03&0&362\\
		2016&73.38&106.75&2.19&4.44&2&452\\
		\bottomrule
	\end{tabular}
\end{table}

In addition, Tables \ref{tb3} and \ref{tb4} record the individual observations of reported and settled claim counts with corresponding frequencies. For instance, in 2012, there were 2 records of reported claims count of 19, and 6 records of cases where no claims were settled.

\begin{table}
	\centering
	\caption{Restructured frequency of reported claims}
	\begin{tabular}{llllll}
		\toprule
		\multirow{2}{*}{Number of reported claims}	& \multicolumn{5}{c}{Frequency of counts by year}\\\cmidrule{2-6}
		& 2012 & 2013& 2014& 2015& 2016\\\hline
		0 & 6 & 5 & 4 & 4 & 3 \\
		2 & 0 & 0 & 0 & 0 & 2 \\
		5 & 0 & 1 & 0 & 0 & 0 \\
		10 & 1 & 0 & 0 & 0 & 0 \\
		19 & 2 & 0 & 0 & 0 & 0 \\
		\vdots & \vdots & \vdots & \vdots & \vdots & \vdots \\
		1520 & 0 & 0 & 0 & 1 & 0 \\
		1533 & 0 & 1 & 0 & 0 & 0 \\
		2053 & 0 & 2 & 0 & 0 & 0 \\
		2151 & 0 & 0 & 0 & 1 & 1 \\
		9902 & 0 & 0 & 0 & 0 & 1 \\
		\textbf{Total} & \textbf{29} & \textbf{29} & \textbf{29} & \textbf{29} & \textbf{29} \\
		\bottomrule
	\end{tabular}
	\label{tb3}
\end{table}

\begin{table}
	\centering
	\caption{Restructured frequency of settled claims}
	\begin{tabular}{llllll}
		\toprule
		\multirow{2}{*}{Number of settled claims}	& \multicolumn{5}{c}{Frequency of counts by year}\\\cmidrule{2-6}
		& 2012 & 2013& 2014& 2015& 2016\\\hline
		0 & 6 & 5 & 4 & 4 & 3 \\
		1 & 1 & 1 & 0 & 0 & 0 \\
		2 & 0 & 1 & 0 & 1 & 2 \\
		3 & 1 & 0 & 0 & 0 & 0 \\
		7 & 1 & 0 & 0 & 0 & 1 \\
		\vdots & \vdots & \vdots & \vdots & \vdots & \vdots \\
		180 & 1 & 0 & 0 & 0 & 1 \\
		307 & 0 & 1 & 1 & 1 & 0 \\
		362 & 1 & 0 & 0 & 1 & 1 \\
		392 & 1 & 0 & 0 & 0 & 0 \\
		452 & 1 & 0 & 0 & 0 & 1 \\
		\textbf{Total} & \textbf{29} & \textbf{29} & \textbf{29} & \textbf{29} & \textbf{29} \\
		\bottomrule
		
	\end{tabular}
	\label{tb4}
\end{table}

It should be noted that, the respective columns for the count frequencies for both reported and settled claims sum up to 29. Thus, each of them tallies with the total number of non-life insurers from whom records are gathered by the National Insurance Commission. Lastly, non-settlement of reported claims, among other reasons, may result from eligibility of reported interest, proximity of cause of insured event, and non-compliance with coverage provisions of the insurance policy. 

\subsection{Model Fitting and Discussion of Results}
This section presents the outcomes of the model fitting methods discussed in Section \ref{sec3} on the claims data from the preceding section. 

The parameter estimates are obtained through maximum likelihood method. The maximum likelihood estimation of the negative binomial and discrete generalised Pareto parameters are performed in R. The negative binomial parameters are estimated using the \emph{mle} function and its standard arguments in the \emph{fitdistrplus} package. However, to the best of the authors'  knowledge, no statistical package exists for estimating the parameters of the discrete generalised Pareto in R. Therefore, the authors wrote an R-function for estimating the parameters of the discrete generalised Pareto distribution, using the algorithm A1-A3, and it is available upon request. In addition, the selection criteria for model comparison are presented for the individual years and the aggregated claims data for the five-year period.

\subsubsection{Parameter Estimation for Yearly Data}
\begin{table}[htp!]
	\centering\
	\caption{Parameter estimates from negative binomial distribution}
	\label{tab: Negative Binomialest}
	\begin{tabular}{lcccccc}
		\toprule
		Variable&Estimator &2012&2013&2014&2015&2016\\
		\hline
		\textbf{	Reported 	Claims}&$\hat{r}$&0.4915&0.6874&1.0370&0.8603&0.4518\\
		&s.e &(0.1234)&(0.1661)&(0.2605)&(0.2121)&(0.1052)\\
		[2ex]
		&$\hat{m}$&296.8624&542.4572&327.0975&396.9273&783.7698\\
		&s.e&(84.7788)&(130.9043)&(64.3623)&(85.7927)&(233.2420)\\
		[2ex]\hline
		\textbf{	Settled Claims}&$\hat{r}$&0.4225&0.6735&1.1312&0.7453&0.7815\\
		&s.e&(0.1063)&(0.1764)&(0.3002)&(0.1878)&(0.1912)\\
		[2ex]
		&$\hat{m}$&88.3519&51.5716&60.2751&71.7227&81.8596\\
		&s.e&(26.7214)&(12.4032)&(11.2194)&(16.3761)&(18.2495)\\\hline
	\end{tabular}
\end{table}

Table \ref{tab: Negative Binomialest} shows the parameters of the negative binomial distribution estimated using an alternative parametrisation given by $X \sim Negative Binomial(r, m/(m + r),$ where $\hat{r}$ and $\hat{m}$ represent the mean and dispersion parameters respectively. The standard errors are placed in parenthesis.

\begin{table}[htp!]
	\centering\
	\caption{Parameter estimates from discrete generalised Pareto distribution}
	\label{tab:  Discrete Generalised Paretoest}
	\begin{tabular}{lcccccc}
		\toprule
		Variable&Estimator &2012&2013&2014&2015&2016\\
		\hline
		\textbf{Reported Claims}&$\hat{\mu}$&7.3983&4.8507&5.8075&6.2320&2.2437\\
		&s.e&(3.4175)&(3.9254)&(3.8876)&(5.6555)&(2.2000)\\
		[2ex]
		&$\hat{\alpha}$&0.0128&0.0044&0.0020&0.0013&0.0102\\
		&s.e&(0.0104)&(0.0011)&(0.0019)&(0.0010)&(0.0017)\\
		[2ex]
		&$\hat{\lambda}$&3.3700&2.7590&2.8036&2.4566&4.9025\\
		&s.e&(0.1318)&(0.8976)&(0.0106)&(0.0829)&(2.3947)\\
		[2ex]\hline
		\textbf{	Settled Claims}&$\hat{\mu}$&3.5866&3.3776&4.6485&3.6443&3.0205\\
		&s.e&(1.0333)&(0.1015)&(3.4941)&(0.6387)&(2.5885)\\
		[2ex]
		&$\hat{\alpha}$&3.4157&2.9258&2.0675&2.3356&0.1121\\
		&s.e&(0.0532)&(0.0200)&(0.0235)&(0.0064)&(0.0311)\\
		[2ex]
		&$\hat{\lambda}$&3.4015&9.4855&3.1532&3.1127&2.8593\\
		&s.e&(0.9538)&(3.0089)&(0.2831)&(1.5265)&(0.9741)\\\bottomrule
	\end{tabular}
\end{table}

Also, Table \ref{tab:  Discrete Generalised Paretoest} presents estimates from the estimators, $\hat{\mu}$, $\hat{\alpha}$ and $\hat{\lambda},$ representing the estimated  discrete generalised Pareto distribution's location, shape and scale parameters respectively. The bootstrap standard errors are placed in parenthesis. 

\begin{table}[htp!]
	\centering\
	\caption{AIC statistics for negative binomial and  discrete generalised Pareto distribution}
	\label{tab: aicdisagg}
	\begin{tabular}{lccccc}
		\toprule
		Distribution&2012&2013&2014&2015&2016\\
		\hline
		\textbf{Reported claims}&&&&&\\
		Negative binomial &329.0927&366.1941&343.5535&352.7973&372.5812\\
		Discrete generalised Pareto &55.9362&55.9999&55.9998&55.9999&55.9972\\
		[2ex]
		\textbf{Settled claims}&&&&&\\
		Negative binomial &274.8660&259.0694&269.3573&277.0937&284.2884\\
		Discrete generalised Pareto &57.9813&57.9958&57.9476&57.9653&57.8863\\
		\bottomrule
	\end{tabular}%
\end{table}

\begin{table}[htp!]
	\centering
	\caption{BIC statistics for negative binomial and  discrete generalised Pareto distribution}
	\label{tab: bicdisagg}
	\begin{tabular}{lccccc}
		\toprule
		Distribution &2012&2013&2014&2015&2016\\
		\hline
		\textbf{Reported claims}&&&&&\\
		Negative binomial &331.5304&368.6319&345.9913&355.2351&375.0189\\
		Discrete generalised Pareto &57.1299&57.1936&57.1935&57.1936&57.1909\\
		[2ex]
		\textbf{Settled claims}&&&&&\\
		Negative binomial &277.3822&261.5856&271.8735&279.6099&286.8046\\
		Discrete generalised Pareto &59.1750&59.1895&59.1413&59.1590&59.0800\\
		\bottomrule
	\end{tabular}%
\end{table}

In terms of reported claims count, Table \ref{tab: aicdisagg} shows that the discrete generalised Pareto model presents smaller AIC and BIC values, in comparison with the negative binomial model. The observation is consistent across the fiscal timelines under consideration.  In addition, for the settled claim counts, the  discrete generalised Pareto distribution is preferred as it exhibits smaller AIC and BIC values throughout the period as shown in Table \ref{tab: bicdisagg}. Therefore, the  discrete generalised Pareto distribution is recommended as it provides a better fit to both classes of the non-life insurance claims data.

\subsubsection{Parameter Estimation for Aggregated Data}
This section presents the results of fitting the negative binomial and  discrete generalised Pareto distributions to the aggregated 5-year count data on reported and settled claims. The results of the parameter estimation for negative binomial and discrete generalised Pareto distributions are presented in Tables \ref{tab: Negative Binomialestagg} and \ref{tab:  Discrete Generalised Paretoest2} respectively. 

\begin{table}[htp!]
	\centering
	\caption{Parameter estimates from negative binomial distribution}
	\label{tab: Negative Binomialestagg}
	\begin{tabular}{lcc}
		\toprule
		Variable&\multicolumn{2}{c}{Estimator (s.e)}\\
		& $\hat{r}$& $\hat{m}$\\
		\hline
		Reported claims &0.6538 (0.0710) &477.0055 (53.2312)\\
		[2ex]
		
		
		Settled claims &0.8596 (0.0978) &74.7975 (7.3175) \\
		
		\bottomrule
	\end{tabular}
	
\end{table}

\begin{table}[htp!]
	\centering
	\caption{Parameter estimates from  discrete generalised Pareto distribution}
	\label{tab:  Discrete Generalised Paretoest2}
	\begin{tabular}{lccc}
		\toprule
		Variable&\multicolumn{3}{c}{Estimator (s.e)}\\
		& $\hat{\mu}$& $\hat{\alpha}$ & $\hat{\lambda}$\\
		\hline
		Reported claims &2.3354 (1.9731) &0.0017 (0.0012)& 2.8171 (2.0240) \\[2ex]
		
		Settled claims &1.9089 (1.6214) &0.0077 (0.0015)&1.9333 (1.4524)\\
		
		\bottomrule
	\end{tabular}
	
\end{table}

\begin{table}[htp!]
	\centering
	\caption{Information criteria for model fitting}
	\begin{tabular}{lcc}
		\toprule
		Model&AIC &BIC\\
		\hline
		\textbf{Reported claims}& &\\
		Negative binomial distribution&1750.1770& 1755.8010\\
		Discrete generalised Pareto distribution&249.9887 & 251.1824\\[2ex]
		
		\textbf{Settled claims}& & \\
		Negative binomial distribution&1311.177& 1316.801\\
		Discrete generalised Pareto distribution&251.9658& 253.1595\\
		\bottomrule
	\end{tabular}
	\label{AIC}
\end{table}

In comparing the negative binomial and discrete generalised Pareto distributions, Table \ref{AIC} shows the AIC and BIC values for the fit of the two probability distributions. It is obvious that discrete generalised Pareto distribution model produces smaller AIC and BIC values for the aggregate reported claim counts. Also, regarding the aggregate settled claim counts, smaller AIC and BIC values are produced by the  discrete generalised Pareto distribution. Therefore, in alignment with the year-based modelling conclusion, the  discrete generalised Pareto distribution is recommended, as it provides a better fit to both classes of yearly data and the aggregated non-life insurance claims data.


\section{Conclusion}\label{sec5}
The study demonstrates that non-life insurance claims frequency can be described by the three-parameter discrete generalised Pareto distribution. Relative to the negative binomial, the discrete generalised Pareto provides a better fit to the distribution of non-life claims data regarding reported and settled counts, as evident in both cases of disaggregated and aggregated year periods. Additionally, profiling non-life insurance datasets by the number of claims and corresponding frequency of counts was shown to provide a data structure for efficient statistical modelling. Further, in evaluating initial estimators ($\alpha_0$, $\lambda_0$, $\mu_0$) for determining the discrete generalised Pareto parameters, the constant unit increment in the $\mu-$frequency and $(\mu + 1)-$frequency methods of \citet{Prieto2014} were modified to the  $\mu$ and $\mu + \epsilon,$ $\epsilon > 0$, where $\mu + \epsilon$ is the smallest observation greater than $\mu$ (the minimum value). This frequency method, $\mu$ and $\mu + \epsilon,$ extends the application of $\mu$ and $(\mu + 1)$ frequency method to practical count data exhibiting varying observational intervals.  


\bibliographystyle{apalike}

\end{document}